\documentclass[12pt,a4paper]{article}

\usepackage{moreverb}

\usepackage[colorlinks,bookmarksopen,bookmarksnumbered,citecolor=red,urlcolor=red]{hyperref}

\usepackage{caption}
\usepackage{subcaption}
\usepackage{wrapfig}
\usepackage[english]{babel}
\usepackage[T1]{fontenc}
\usepackage[ansinew]{inputenc}
\usepackage[pdftex]{graphicx}
\usepackage{hyperref}

\newcommand\BibTeX{{\rmfamily B\kern-.05em \textsc{i\kern-.025em b}\kern-.08em
T\kern-.1667em\lower.7ex\hbox{E}\kern-.125emX}}

\usepackage{fancyhdr}
\begin{document}

\begin{center}
{\LARGE \bf Design and Optimisation of an Efficient HDF5 I/O Kernel for Massive Parallel Fluid Flow Simulations} \vspace{0.7cm}

{\large \em Christoph Ertl$^1$, J\'{e}r\^{o}me Frisch$^2$, and Ralf-Peter Mundani$^1$} \vspace{0.1cm}

{\em \{christoph.ertl, frisch, mundani\} @ tum.de} \vspace{0.1cm}

{\em $^1$ Technische Universit\"at M\"unchen, Arcisstrasse 21, 80290 M\"unchen, Germany}\\
{\em $^2$ RWTH Aachen University, Mathieustrasse 30, 52074 Aachen, Germany}

\end{center}

\begin{abstract}
More and more massive parallel codes running on several hundreds of thousands of cores enter the computational science and engineering domain, allowing high-fidelity computations on up to trillions of unknowns for very detailed analyses of the underlying problems. During such runs, typically gigabytes of data are being produced, hindering both efficient storage and (interactive) data exploration. Here, advanced approaches based on inherently distributed data formats such as HDF5 become necessary in order to avoid long latencies when storing the data and to support fast (random) access when retrieving the data for visual processing. Avoiding file locking and using collective buffering, write bandwidths to a single file close to the theoretical peak on a modern supercomputing cluster were achieved. The structure of the output file supports a very fast interactive visualisation and introduces additional steering functionality.
\end{abstract}

\vspace{0.2in}

\section{Introduction}
According to the U.S.\ President's Council of Advisors on Science and Technology \emph{`high-performance computing must now assume a broader meaning, encompassing not only flops, but also the ability, for example, to efficiently manipulate vast and rapidly increasing quantities of both numerical and non-numerical data'} \cite{bib:kalilmiller2015}. Latest advances in hardware led to high-performance computing (HPC) systems consisting of hundreds of thousands to millions of cores that exhibit petaflop performance for high-fidelity applications. Designing massive parallel codes that can utilise such systems is a challenging task itself,  being able to handle the inherent huge data advent -- easily exceeding tons of gigabytes per computational step -- is yet another one. This raises the necessity for sophisticated concepts to store and interact with the computed data---even in real-time.

Within the authors' research, they have developed a computational fluid dynamics (CFD) code for multi-scale, multi-physics problems arising in the field of computational engineering with special emphasis on thermal comfort assessment. This code has been successfully deployed on up to a total of 140,000 processes on two of Germany's three top-ranked supercomputers, namely SuperMUC (based on System x\textsuperscript{\textregistered} iDataPlex\textsuperscript{\textregistered} dx360 M4 compute nodes) installed at Leibniz Supercomputing Centre and JuQueen (Blue Gene/Q system) installed at J\"ulich Supercomputing Centre. Furthermore, performance measurements were done using up to 32,000 processes on Shaheen (Blue Gene/P system) installed at KAUST Supercomputing Laboratory. Core part of the code is a hierarchical data structure consisting of logical and computational grids following a space-tree based \cite{bib:babufrmu2002} spatial partitioning. In conjunction with a so-called neighbourhood server, a topological repository storing which computational grid resides on which process, this structure fosters distributed computing and supports efficient numerical solvers as well as dynamic load balancing strategies.

Furthermore, the structure also supports in-situ monitoring of the computed data, in other words, users have the possibility to retrieve computation results already during runtime for visual exploration. A sliding window called approach allows to interactively select any region of the computational domain, whereupon the size of the window defines the corresponding level-of-detail, thus keeping the total amount of data to be transmitted between the CFD code and the user constant in order not to exceed given bandwidth limitations \cite{bib:mufrvara2015}. Hence, a window covering the entire domain allows for a qualitative evaluation of the global flow field while smaller windows -- cf.\ a zooming into the data -- reveal more and more details for quantitative assessments. Even for huge domains with trillions ($10^{12}$) of unknowns, the sliding window approach is advantageous as only small parts of the data need to be processed for visual display. For further research regarding actual HPC in-situ visualisations the reader is advised to Rivi et al.\ \cite{bib:ricamusl2012}.

Nevertheless, at certain time steps the data will be written to a storage system either for checkpointing (fault tolerance) purposes or an offline post-processing of the computation results. During this time, all processes cannot continue with their computations and have to wait until the data dump has been finished, thus slowing down the entire execution. Finally, the user is left with tons of sequentially ordered data that -- due to its mere size -- forbid any efficient access or treatment within subordinated post-processing tools. In order to tackle this problem, an I/O kernel based on HDF5 has been implemented that supports parallel I/O functionality to speed up the write operations of the checkpointing, minimising the impact on the overall execution time of the code. 

The main characteristic of this approach is the mapping of a rather complex, but highly efficient data structure in terms of parallelisation and selective visualisation to an I/O kernel based on HDF5. The authors were able to show competitive write bandwidths on top tier machines, even when comparing with a state-of-the-art I/O kernel employing a comparable lighter data structure. Including not only raw data values, but also detailed information about the domain topology at every checkpoint, the offline structure enables very fast restarts, without the need to reconstruct the domain. Additionally, it allows to utilise the sliding window approach even on offline data. Hence, users can switch between online (present) and offline (past) data for visual exploration---practically they can reverse in time. Such a time reversal further provides the possibility to modify a scenario at any point in time and re-compute it with altered settings in case of undesired results or effects, thus opening the door for computational steering or interactive computing applications.

This paper is based on Ertl et. al. \cite{bib:erfrmu2016}, but the current paper includes the following additional research: A more thorough look on the time reversible steering concept with supporting simulation scenarios. Furthermore, the I/O kernel was deployed and tested on SuperMUC. 
 
The remainder of this paper is organised as follows. In the next section, the basic mathematical concepts as well as the data structure are introduced, including some runtime and speed-up measurements done on SuperMUC, JuQueen and Shaheen. The design and implementation of the I/O kernel using HDF5 is addressed in section three. Section four gives a more detailed introduction into the time-reversible steering concept based on the kernel's functionality. The following section five presents write speed measurements of full sized runs on JuQueen and SuperMUC. Finally, section six is closing with a short summary.

\section{Fluid Flow Simulations}
In this section, the foundations of the CFD code \emph{mpfluid} including all mathematical concepts are presented. Main contribution is a hierarchical data structure inherently supporting distributed computing and allowing for in-situ data exploration already during runtime. This follows the elucidations of \cite{bib:frmuratr2015}.

\subsection{Mathematical Modelling}
The implementation of the CFD kernel is based on the Navier--Stokes equations which can be derived from the conservation of mass, momentum, and energy principles \cite{ bib:ferzperic2002} \cite{bib:hirsch2007}.

The governing equations for incompressible Newtonian fluid flows consist of the continuity equation and the momentum equations. The continuity equation can be written in differential form as
\begin{equation}
\label{eqn:continuity}
\nabla\cdot\vec{u} = 0\,,
\end{equation}
where $\vec{u}$ describes the velocity vector of the flow field. For a divergence free vector field, the continuity equation has to be satisfied at every time step in the entire domain.

The momentum equations for every direction $i \in \{1, 2, 3\}$ can be written in differential form as
\begin{equation}
\label{eqn:momentum}
\frac{\partial\rho_{\infty}u_i}{\partial t} + \nabla\cdot(\rho_{\infty}u_i\vec{u}) = \nabla\cdot(\mu
\nabla u_i) - \nabla\cdot(p\vec{e}_i) + b_i\,,
\end{equation}
where $t$ represents the time, $\rho_{\infty}$ the density of the fluid (assumed constant over the entire domain), $u_i$ the velocity in direction $i$, $\mu$ the dynamic viscosity, $p$ the pressure, $b_i$ some interior body forces in direction $i$, and $\vec{e}_i$ the unit vector in direction $i$.

In order to include thermal effects such as buoyancy, the last part $b_i$ on the right-hand side in (\ref{eqn:momentum}) must be replaced by $\rho_{\infty}\cdot\beta\cdot(T-T_{\infty})g_i$ to couple effects of the temperature field to the momentum equations, commonly known as the Boussinesq approximation, see Lienhard and Lienhard \cite{bib:lienhard2011} for instance. Here, $\beta$ describes the thermal expansion coefficient of the fluid, $T$ the temperature, $T_{\infty}$ the temperature of the undisturbed fluid at rest, and $g_i$ the gravitational force in direction $i$. Finally, for modelling the thermal heat transport, the energy equation (adapted from a generic convection-diffusion equation to a heat transfer problem) can be written in differential form as
\begin{equation}
\label{eqn:energy}
\frac{\partial T}{\partial t} + \nabla\cdot(T\vec{u}) - \nabla\cdot(\alpha\nabla T) - \frac{q_{int}}{\rho_{\infty}\cdot c_p} = 0\,,
\end{equation}
where $\alpha = k/(\rho_{\infty}\cdot c_p)$ represents the heat diffusion coefficient, $k$ the heat conduction coefficient, $c_p$ the specific heat capacity at constant pressure, and $q_{int}$ the internal heat generation.

As pressure $p$ is solely contained in its gradient form  $\nabla p$ in (\ref{eqn:momentum}), some pressure correction methods as proposed by Harlow and Welch \cite{bib:harlwelch1965} or Chorin \cite{bib:chorin1968} have to be applied. In the present approach, the so-called fractional step or projection method introduced by Chorin that iterates between the velocity and pressure fields is used, where the latter one acts as correction to the velocity field in every time step to fulfil (\ref{eqn:continuity}).
Choosing an explicit Euler time discretisation for the temporal derivatives $\partial/\partial t$, one is eventually left with a Poisson equation for the pressure that has to be solved in every time step and also determines the computationally most complex part. For the spatial discretisation a finite volume method is applied that -- due to the block substructuring of the domain into regular Cartesian grids -- locally degenerates into finite differences and, thus, favours fast computations based on standard stencil operators.

In the next part, the data structure together with a multigrid-like solver -- directly derived from the data structure's exchange routines -- for the solution of the pressure Poisson equation will now be introduced.

\subsection{Data Structure}

\begin{wrapfigure}[29]{r}{0.50\textwidth}
\vspace{-20pt}
\centering
\includegraphics[width=0.30\textwidth]{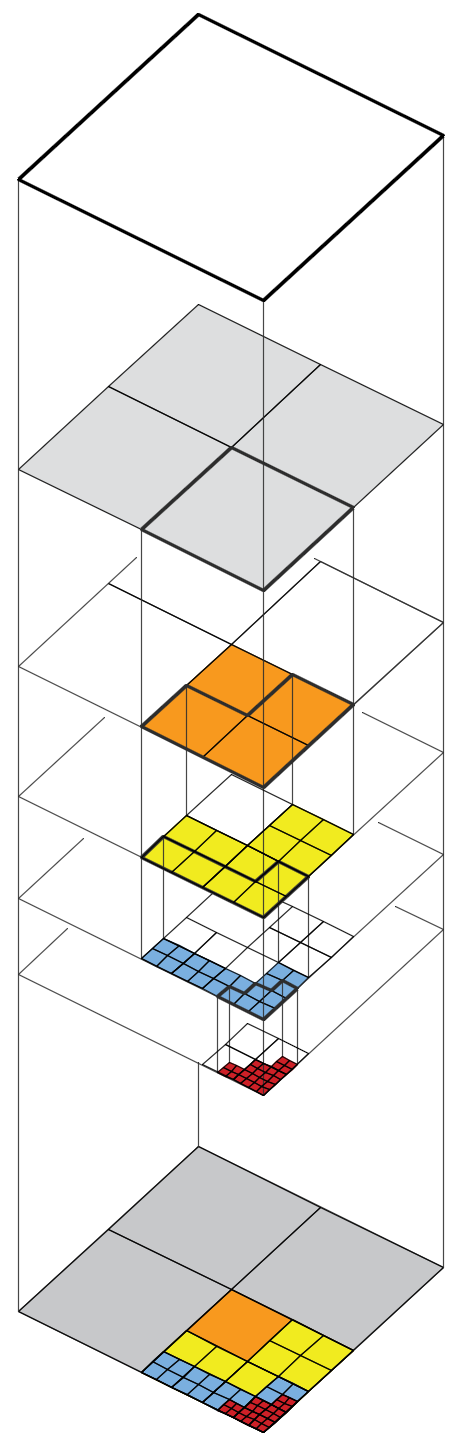}
	\captionsetup{width=0.50\textwidth}
  \caption{Example 2D data structure, adaptively refined up to depth 5}
  \label{fig:structure}
\end{wrapfigure}

The data structure follows the general idea of space-trees (with quadtrees as 2D and octrees as 3D representatives) for a spatial partitioning. Starting from a single root cell on depth $0$, each cell is further subdivided by $r_x \times r_y \times r_z$ cells until a predefined depth $d_{max}$ has been reached. This hierarchy of grids defines the logical part of the structure -- also called logical grid or l-grid -- and plays a vital role when retrieving any hierarchic grid information. For computation purposes, now every cell of this logical grid links to a data grid of size $s_x \times s_y \times s_z$ that stores all necessary variables such as velocities, pressure, or temperature values. Furthermore, each data grid -- also referred to as d-grid -- is surrounded by a halo (currently of size one) for the proper data exchange between d-grid boundaries. This completes the data structure which is composed of block-structured, non-overlapping, orthogonal, regular, hierarchical grids. An example 2D data structure is illustrated in Fig. \ref{fig:structure}. The root grid on top is successively refined up to depth 5, using a bisection in both dimensions, that is $r_x = r_y = 2$. In addition, the example shows that the data structure also supports adaptive subdivision of grids, such that it is possible to represent regions of interest with a finer resolution. 

To keep track about the distribution of d-grids to (MPI) processes, a dedicated process called neighbourhood server was implemented. This server stores the entire logical structure, the l-grids, in order to answer topological queries, while all computational processes solely store the d-grids assigned to them. The assignment per se follows a space-filling Lebesgue curve that has proven to preserve neighbouring relations, thus reducing the necessary communication overhead. For a ghost layer update, any computational process queries the neighbourhood server by its own d-grid IDs in order to obtain all neighbouring d-grid IDs and their physical residence (i.\,e.\ MPI ranks). Afterwards, the process can launch an inter-grid data transfer (communication phase) which is strictly separated from the computation phase.

\begin{figure}[!t]
  \begin{subfigure}[b]{0.65\textwidth}
    \includegraphics[width=\textwidth]{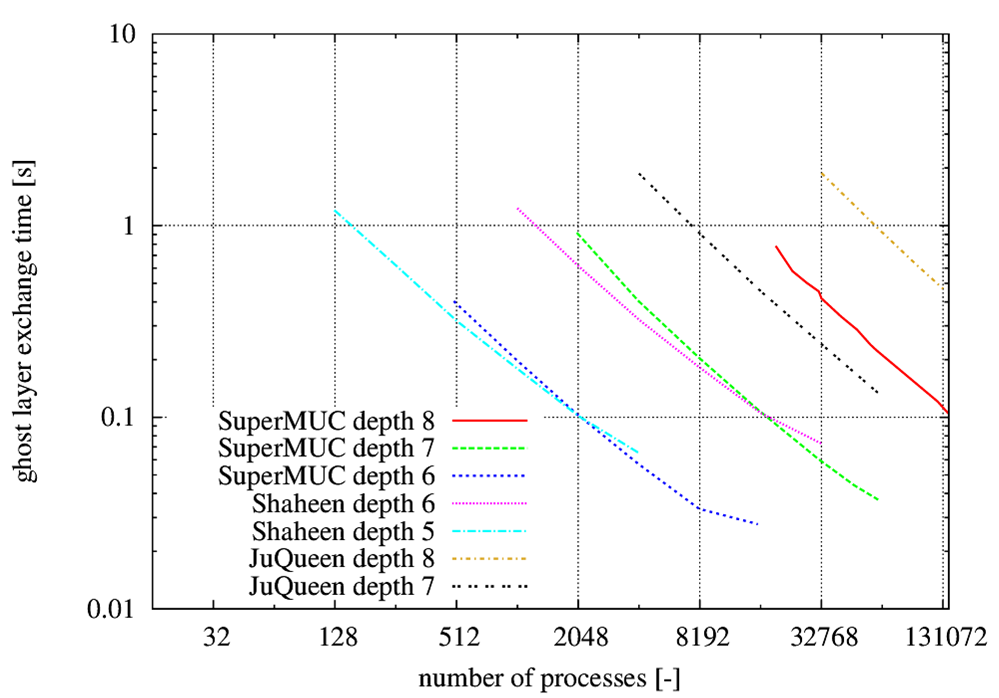}
    \caption{Total ghost layer exchange times for different amounts of processes}
    \label{fig:halo_update}
  \end{subfigure}
  \hfill
  \begin{subfigure}[b]{0.65\textwidth}
    \includegraphics[width=\textwidth]{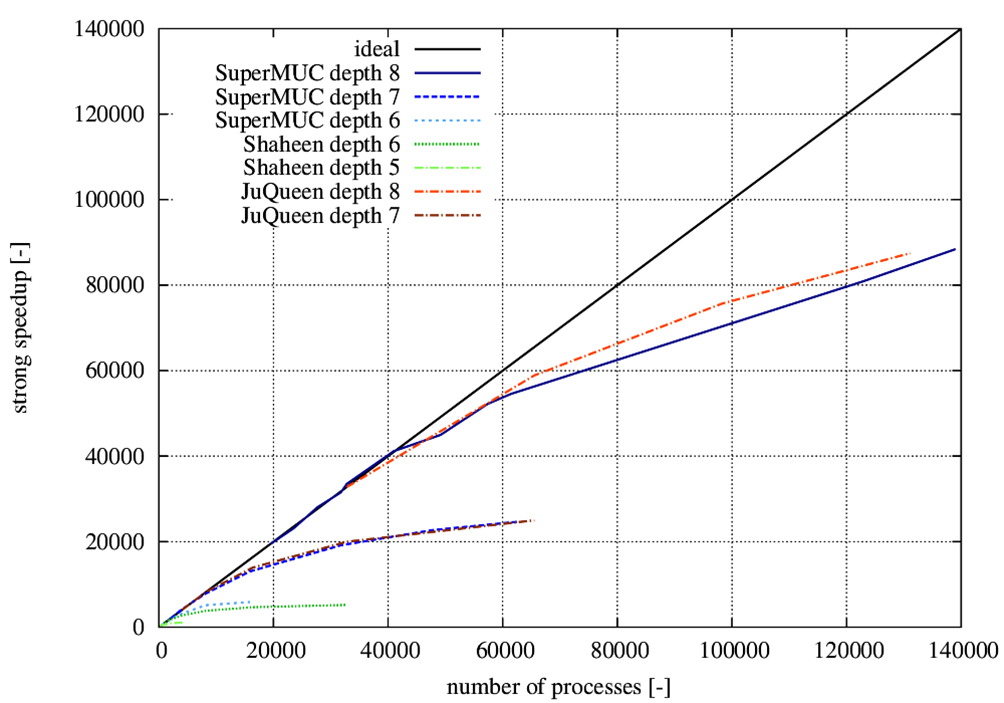}
    \caption{Strong speed-up values}
    \label{fig:speedup}
  \end{subfigure}
    \hfill
  \centering
    \begin{subfigure}[b]{0.65\textwidth}
    \includegraphics[width=\textwidth]{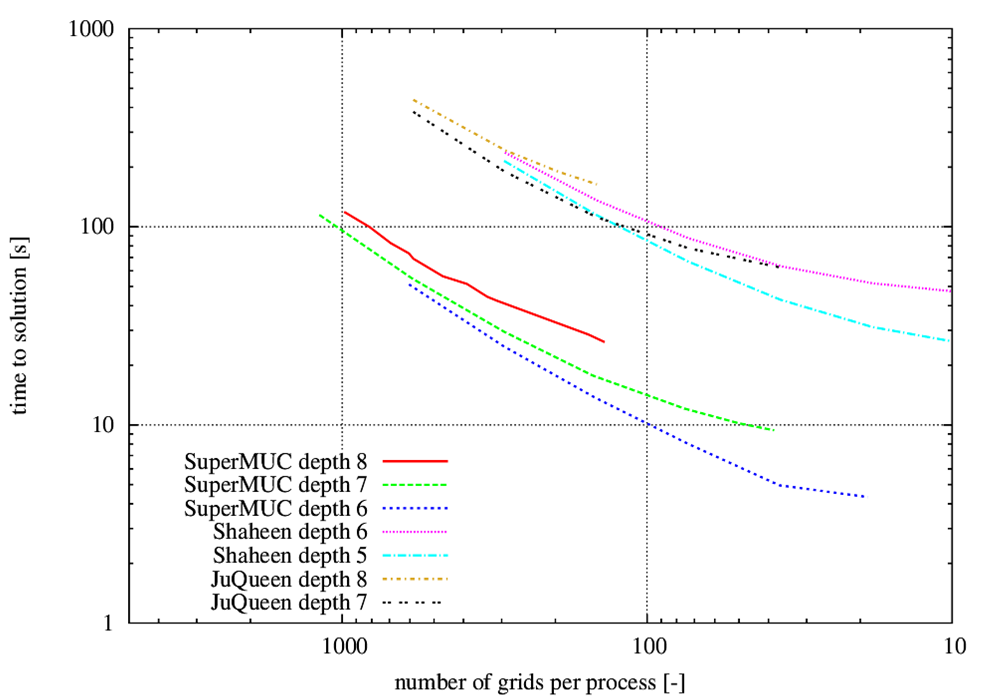}
    \caption{Time-to-solution (one full time step) plotted against the number of d-grids per processes}
    \label{fig:timetosolution}
  \end{subfigure}
  \caption{Characteristic plots for different resolutions up to $4096 \times 4096 \times 4096$ (depth 8) with approx.\ 707 billion unknowns on different HPC systems}
\end{figure}

The communication phase consists of three sequential steps as described in \cite{bib:frmura2011}. First of all, in a bottom-up step all d-grids that have not been updated yet during the computation phase are set to the averaged values of their corresponding child d-grids. In a second, horizontal step all adjacent d-grids update their ghost layers before in a last, top-down step all resulting ghost layers of d-grids on different levels (due to an adaptive grid refinement) are set properly. Here, the l-grid management must take care about flux conservation across d-grid boundaries in order to guarantee data integrity and consistency. Whereas the communication phase is not very time consuming -- a full update for a domain with resolution $4096 \times 4096 \times 4096$ resulting to more than 700 billion unknowns to be exchanged takes around 0.1\,s on 140,000 cores on SuperMUC, see Fig.~\ref{fig:halo_update} -- according to \cite{bib:Frisch2014} the parallel code spends more than 90\,\% of the time in the computation phase for solving the pressure Poisson equation.

To solve the pressure equation, a parallel multigrid-like solver following the ideas of Brandt \cite{bib:brandt1977} for solving elliptic partial differential equations was implemented. Multigrid-like, because it utilises the above communication schema -- precisely the bottom-up and top-down update steps -- as restriction and prolongation operators for setting up a cell-centred multigrid method, thus making use of the data structure's superior parallel performance and scalability properties. Nevertheless, the multigrid-like solver exhibits convergence instabilities for certain scenarios (in case of adaptive refinement, e.\,g.) which can be handled by different smoothing strategies such as doubling the amount of pre- and post-smoothing steps on coarser resolutions. Details about all performed analyses and comparisons can be found in \cite{bib:frmura2013}. Fig.~\ref{fig:speedup} and Fig.~\ref{fig:timetosolution} show the obtained strong speed-up and time-to-solution values of the multigrid-like solver for different domains on different architectures.

\subsection{Sliding Window Visualisation}
To reduce the data transfer between the CFD code running on an HPC system -- called back end -- and the user application for visual display -- called front end -- the previous data structure was extended with the idea of the sliding window concept. Main strategy is to select on the back end only subsets of the computed data in order to stay below the available network bandwidth between front and back end. Hence, any user has the possibility to choose a region of interest (the window) which can be moved around the computational domain and increased or decreased in size. The larger the window, the lower will be the density of data points to be considered for the visual display---even when a higher density of data points would be available, depending on the window size every second, third, fourth, and so on data point will be dismissed. This approach allows to catch either global effects of the simulation (large window) or to explore local details (small window) without overloading the network with unnecessary data and, thus, harming the experience of an authentic interactive computing.

\begin{figure}[!t]
\centering
\includegraphics[height=2in]{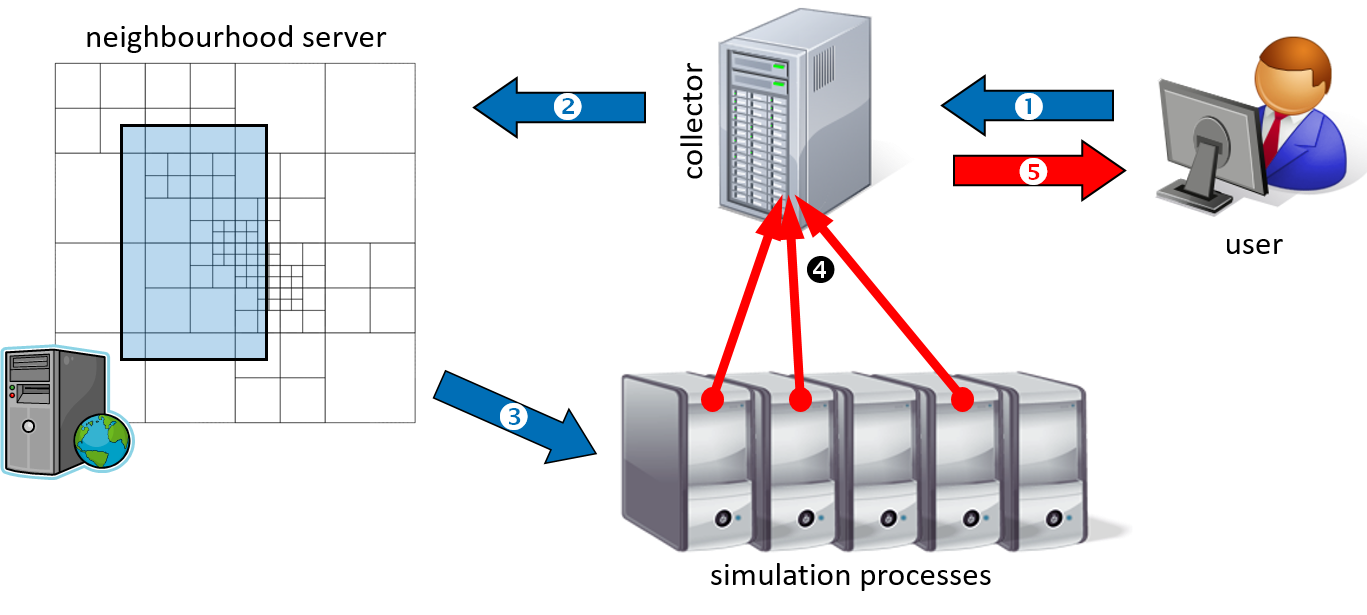}
\caption{Single steps of a full sliding window query: the client issues a request to the collector (1) that forwards it to the neighbourhood server for identifying all relevant d-grids (2) and informs respective computational processes (3) to send the desired information to the collector (4) that finally returns a data stream to the client (5) -- based on \cite{bib:mufrvara2015}}
\label{fig:slidingwindow}
\end{figure}

The sliding window implementation practically consists of two components: a back end collector for gathering the desired data and a front end visualisation and interpreter tool for sending data requests. Therefore, a new process called collector was introduced at the back end, listening for user requests on a standard TCP socket. The collector forwards the query to the neighbourhood server which can easily identify all computational processes storing relevant d-grids intersecting with the window. These processes themselves send all selected data points to the collector which returns a compressed data stream to the front end application. On the front end we use ParaView \cite{bib:paraview} for the visualisation of the fluid flow data. A special ParaView plug-in allows the user to connect to a running simulation, to set all necessary sliding window parameters, and to retrieve the desired data for an interactive visual data exploration. Fig.~\ref{fig:slidingwindow} shows a full pass of a sliding window query.

The Sliding Window approach has shown no significant performance overhead in practice. The data selection process happens on the neighbourhood server, completely independent from any computing process. One additional communication to send selected grids to the collector process is necessary, but is limited by the fixed amount of data transferred by the Sliding Window concept, adjusting the density of data points according to the window size. Aggregating and sending the data over the TCP connection by the collector is again completely independent from the computing processes and thus has no influence on the execution time. 

\section{I/O Kernel}
\label{sec:I/O}
Previously the approach to store and visualise simulation data -- aside from online monitoring -- was based on each computing process writing out the the grid data comprising the domain in the finest resolution available into an individual binary VTK file per time step. Usually this approach performs well, in case of a scarcity of I/O links for example on JuQueen (\ref{subsec:Systems}), having a multitude of processors inquire an I/O operation at the same time leads to severe contention. Furthermore, the sheer number of files generated using one of the top tier machines is hardly manageable by the file system. For post-processing the individual files are appended into a single one per time step, again causing a heavy burden on the file system. Finally, visualising such files, gotten from production sized simulation runs, is not viable in a reasonable amount of time. In order to be able to handle the large amount of data our CFD code produces, a dedicated approach to the code's I/O routines had to be employed.

One major building block of this approach comprises the use of the high-level data format HDF5 (Hierarchical Data Format version 5) \cite{bib:HDF5}. HDF5 is specifically tailored to store large amounts of array based data in a database-like view. The key concept behind HDF5's data model is based on datasets which contain the actual data and groups making up the general structure. Starting from a root group, groups may contain additional groups or datasets themselves, resulting in a hierarchical tree-like structure resembling a Unix file system. Attributes allow to describe the data and can be associated to a group or dataset. The layout of the data is specified by HDF5's storage model, while the HDF5 library takes care of the conversion between the database-like view of the data model and the storage model. The user may be completely oblivious to the way his data looks like on the file system. Considering performance, however, it is necessary to be aware of the layout of the data in an HDF5 file. Considerations concerning this are found in \ref{subsec:Implementation}.

HDF5 is a self-describing format, meaning a file contains information about its structure as well as the used data types. Portability is a huge concern in today's diverse architecture landscape. Different machines and compilers employ different notions of endianess as well as different sized data types, making the transfer and processing of files between machines a non-trivial task. However, during the mapping from storage to data model and vice versa, the HDF5 library accounts for these discrepancies. Using the self-describing information, the data is converted from the source to the target machines' architecture without any attention required from the user.

Additionally, HDF5 provides functionality for distributed memory systems. Parallel HDF5 routines are based on an underlying implementation of MPI-IO, whereas the HDF5 libraries manage the application's I/O calls and in turn utilise MPI-IO's routines, providing easy-to-use parallel I/O functionality.

\subsection{Functionality and Design of the I/O Kernel}
\label{subsec:Design}

The most important aspects of designing an I/O kernel were fixing the desired functionality and subsequently determining on how the tree-structure within the HDF5 file is conceived. Further on, the content and shape of the datasets had to be settled to support the intended functionalities. Due to the fact that structure, functionality, and considerations concerning the implementation strongly influence each other, this process was revised throughout the whole development. The final outcome is based mainly on the following conditions.

To reduce the management effort and the overall load on the file system, a shared file approach is used, in which each participating process reads and writes to a single file. Also, in its current version, this output file supports the complete set of intended functionality since most of these have overlapping requirements. However, this is subject to be revised in future iterations of the kernel to allow users turn off unnecessary functions and, thus, reduce the amount of data in the file.

The following functionality is currently supported by the kernel and the output file:

\begin{itemize}
\item \textit{output} \\
The main purpose of the I/O kernel is to output snapshots of the running simulation at user defined intervals. These snapshots give a complete view of the topological grid structure as well as the computed cell values. The file structure and the I/O routines were conceptualised to achieve a write-out as fast as possible, resulting in a minimal impact on the overall execution time of the CFD code. Apart from good programming practice this is achieved by using hardware specific optimisation. For details see section \ref{subsec:HardwareOp}. 

\item \textit{checkpointing} \\
Using any of the written simulation snapshots, the code is able to recreate the topological grid structure from the HDF5 file and resume computation. This prevents costly data loss after a crash or a power outage for example or allows for splitting time intensive computations into smaller parts, to better utilise the sparse and expensive CPU hours on current high-performance machines. Also, the code's current domain decomposition and distribution strategy is done serially (by the neighbourhood server). To allow for an efficient usage of resources, this step can be prepared on a smaller machine while the simulation run is then started from the HDF5 file. 

\item \textit{offline sliding window} \\
The sliding window concept allows for visualisation during runtime using the neighbourhood servers complete view of the topological grid structure. The neighbourhood server is able to select a subset of simulation data of arbitrary resolution representing the desired section. This allows for an efficient limitation of data to transfer and display. However, this is possible for the currently computed time step only, i.\,e.\ online. Having the hierarchical grid structure also present within the snapshots of the HDF5 file allows for a sliding window scheme in a similar -- offline -- fashion. This enables users to visualise even largest datasets in a quick and efficient manner for every written time step. 

\item \textit{time reversible steering} \\
Being able to follow a running simulation through the online sliding window concept, the I/O kernel further allows influencing a running simulation in a computational steering approach. Changing boundary conditions or issuing refinements and coarsenings at runtime are possible. The ability to visualise previous time steps as well as to recreate the complete topological domain structure and the corresponding data values at any written checkpoint provided by the I/O kernel, enables another dimension in steering capabilities, so called time reversible steering. This enables to go back to a previous time step, load this state and issue steering commands from there, shortening simulation cycles, and allowing an iterative optimisation of simulation parameters. In Section \ref{sec:TRS}, this concept is explained in more detail and shown with possible applications.
\end{itemize}

\begin{wrapfigure}[20]{r}{0.4\textwidth}
\includegraphics[width=0.35\textwidth]{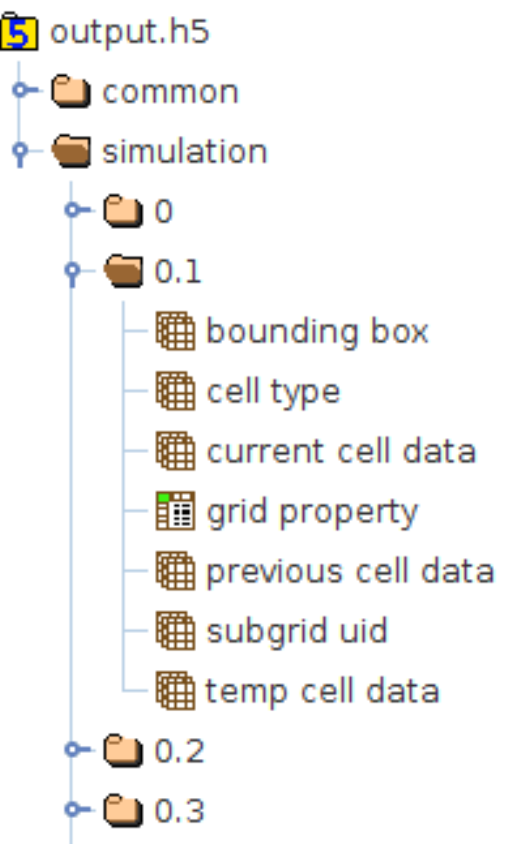}
	\captionsetup{width=0.35\textwidth}
  \caption{Tree-like structure of the HDF5 output file}
  \label{fig:tree}
\end{wrapfigure}

The current structure of the HDF5 file is illustrated in Fig.~\ref{fig:tree}. Below the root group the structure splits into two branches. The \emph{common group} encloses datasets that hold constant information such as the time discretisation step, refinement spacings, or fluid properties. It is therefore only accessed once at generation of the file or start of a simulation run. The \emph{simulation group} holds further groups that enclose the structure information as well as the cell data itself for each time step. These time step groups are named with the elapsed time.  

All datasets in the time step groups follow a simple paradigm. Every row in the two dimensional datasets corresponds to one grid in the code's data structure. Additionally, the row indices are constant across all datasets representing one time step, such that the row index referring to one specific grid is identical in all datasets. 

The grids are ordered by the respective ranks on which they reside in the code. Grids residing on rank 0 occupy row indices starting from 0. Grids on subsequent ranks follow in the same manner. The root grid, that comprises the complete physical domain is by construction always at the top most position on rank 0. A very important consequence of the ordering is that the root grid will always be represented by index 0 in every dataset, providing the necessary starting point to traverse the hierarchical data structure.  

The datasets \textit{grid property}, \textit{subgrid uid}, and \textit{bounding box} make up the topological grid structure of the respective time step. \textit{grid property} stores the Unique Identifier (\textit{UID}) for every grid, encoding the residing rank, a rank unique identifier and its location in the structure. \textit{subgrid uid} contains the \textit{UID}s for grids created by refining the respective grid on the next finer level and \textit{bounding box} encodes the physical extent of each grid. 

The datasets \textit{current cell data}, \textit{previous cell data}, and \textit{temp cell data} contain the data values for every cell of every grid such as velocities, pressure etc. The \textit{cell type} dataset stores the boundary conditions. These last four datasets make up the vast majority of data in the file.

\subsection{Implementation Aspects}
\label{subsec:Implementation}
All HDF5 functionality is encapsulated into one C++ class that each computing process instantiates. The current iteration of the kernel makes no use of any functionality of the neighbourhood server to avoid possible communication bottlenecks. 

The first write creates the file and the aforementioned tree structure while subsequent writes only open the file and add the respective time step group and datasets.

In Parallel HDF5, the group structure as well as every dataset has to be created collectively by all participating ranks, while read and write operations can be carried out individually. To be able to generate the needed datasets, the total amount of grids in the complete domain must be known to all ranks. Additionally, to determine the non overlapping regions of individual ranks in the datasets for their read and writes -- in HDF5 terminology so called hyperslabs -- every rank must be aware of the cumulative amount of grids on previous ranks. This is achieved using a global MPI reduction, summing up all grids, followed by an MPI prefix reduction to determine the amount added by all previous ranks to the global sum.

In the HDF5 storage model, each dataset is represented via a header followed by the actual data in form of a linear array, regardless of its actual dimensionality. The shape of the dataset is defined by the header information. For optimised performance, a one to one mapping of data from the code to the HDF5 file is desirable. For this purpose, a linear write buffer is initialised on each rank in which the grid data is copied. This additional storage requirement effectively cuts the amount of data to be handled by a single process in half. As emphasis is laid on the added performance by this approach, the drawbacks of limiting the amount of data per rank was deemed acceptable.  

Reading and restarting was conceived in a comparable fashion. Each rank reads the \textit{UID}s in the \textit{grid property} dataset to determine their range of grids according to the rank information encoded in the \textit{UID}s. Each rank then generates the respective amount of grids, reads the data from its hyperslab, and copies it to the respective grids. The neighbourhood server registers the topological structure and computation may commence. If restart from an intermediate snapshot is ordered, the I/O kernel creates a new branching file for subsequent write outs. 

The sliding window approach in the code makes use of the ability of the neighbourhood server traversing the logical grid structure from the root downwards to subsequently refined grids. If a sliding window query is send to the neighbourhood server, it successively adds and removes d-grids to a list while traversing the tree until it has found the finest possible resolution fitting into a given limit of bandwidth and visualisation window. The sliding window on top of the HDF5 file uses the same approach of traversing the tree, starting from the root grid at row index 0. This is achieved by assigning the \textit{UID} information of a grid to its respective row index via the \textit{grid property} dataset. Grids on subsequent refinement levels are found via the \textit{subgrid uid} dataset. The routine ends up with a list of indices referring to the grids that fit the determined criteria and allows for a selective visualisation of the corresponding grid data.

\section{Time Reversible Steering (TRS)} \label{sec:TRS}

The visualisation front end can be used to issue commands to the simulation back end and influence the running simulation in a computational steering approach. Possible operations are the ordering of refinements or coarsenings of the simulation space, or the altering of boundary conditions, for example moving geometry or influencing velocity constraints. The new functionality introduced by the HDF5 output file, namely the possibility to visualise and restart from any written time step in a timely fashion, allows for an extension of the classical steering.

\begin{figure}[t]
\centering
\includegraphics[width=0.95\textwidth]{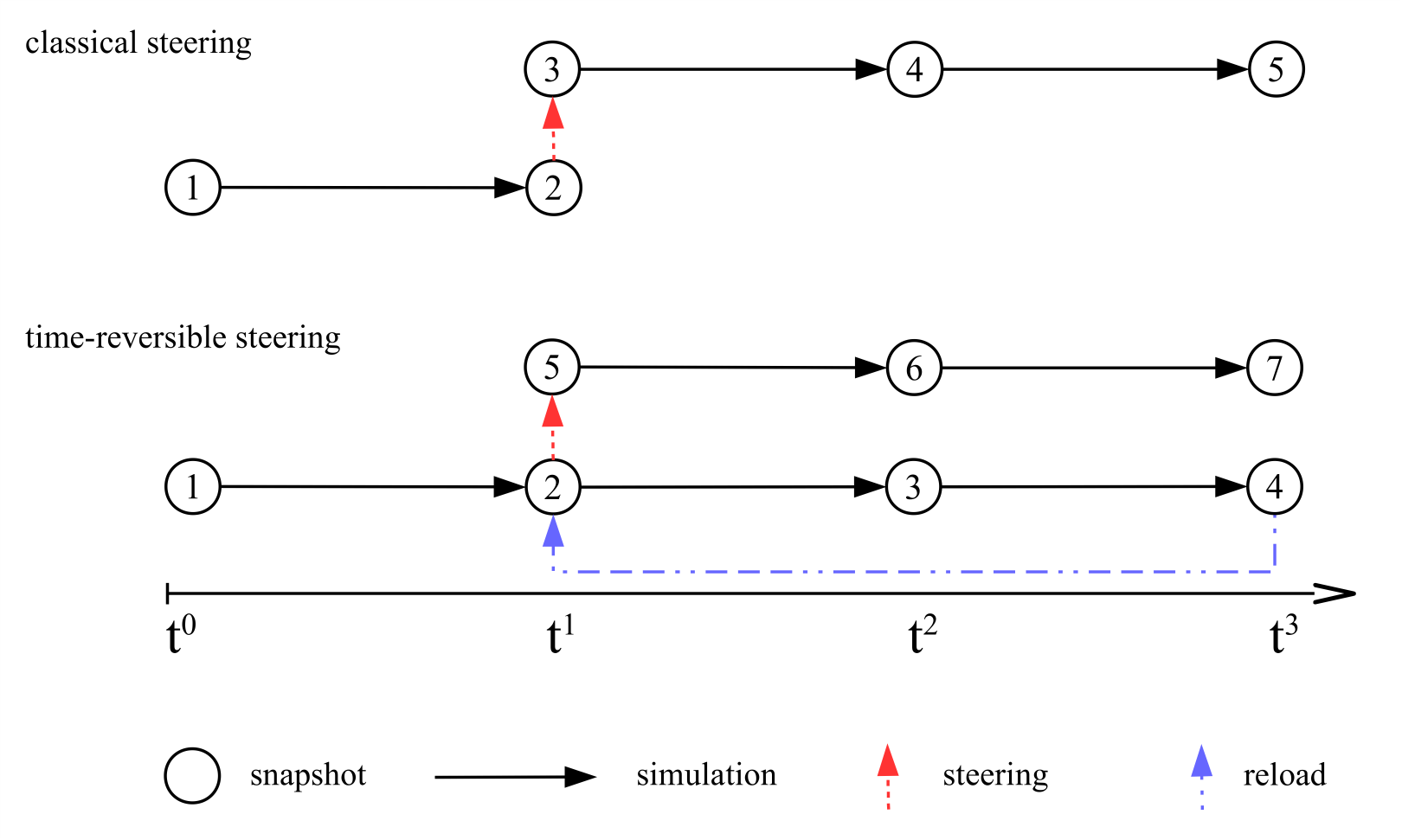}
\caption{Schematic concept of classical and time reversible steering concept. Horizontal shifts signify continuation of the simulation while a vertical shift represents a steering operation. Main characteristic of classical steering at the top is a continuous progression forward in time, whereas time reversible steering allows the reload and altering of previous written checkpoints, leading to branching simulation paths.}
\label{fig:TRS}
\end{figure}

This extension, titled Time Reversible Steering (TRS) alongside the classical steering is schematised in Fig. \ref{fig:TRS}. Simulation snapshots are depicted using circles, the numbers within signify the succession of generation. A horizontal shift signifies continuation of the simulation (black solid arrows), while vertical shifts are used to represent a steering operation (red dashed arrows). The former approach, supported by the back end/front end link and the online visualisation is depicted by a single continuous simulation path. Steering operations influence the current state and the subsequent simulation. The offline visualisation enabled by the HDF5 file grants access to all previously written snapshots, allowing a more thorough understanding of forming flow characteristics. In addition, it is possible to reload every checkpoint from the file (blue dashed and dotted arrows) in rapid fashion. Due to the stored domain topology in the file, it is not necessary to build up and distribute it from scratch. Finally, the issue of steering operations for reloaded checkpoints are carried out in the same fashion as with classical steering. Having two or more possibilities leads to branching simulation paths. 

TRS as well as its predecessor the classical steering are at the moment not suited for full-sized production runs. The overhead of the approach is simply too large to be run on current top tier machines. Nevertheless, the concept's main advantage is the possibility to quickly evaluate the influence of changes to the simulation on a reduced problem, filtering less promising outcomes. This shortens design cycles and leads to more efficient use of the expensive computing time on clusters. To further emphasise on the applicability for the approach two simulation scenarios were conducted and the concept was applied.

\begin{figure}[t]
\centering
\includegraphics[width=\textwidth]{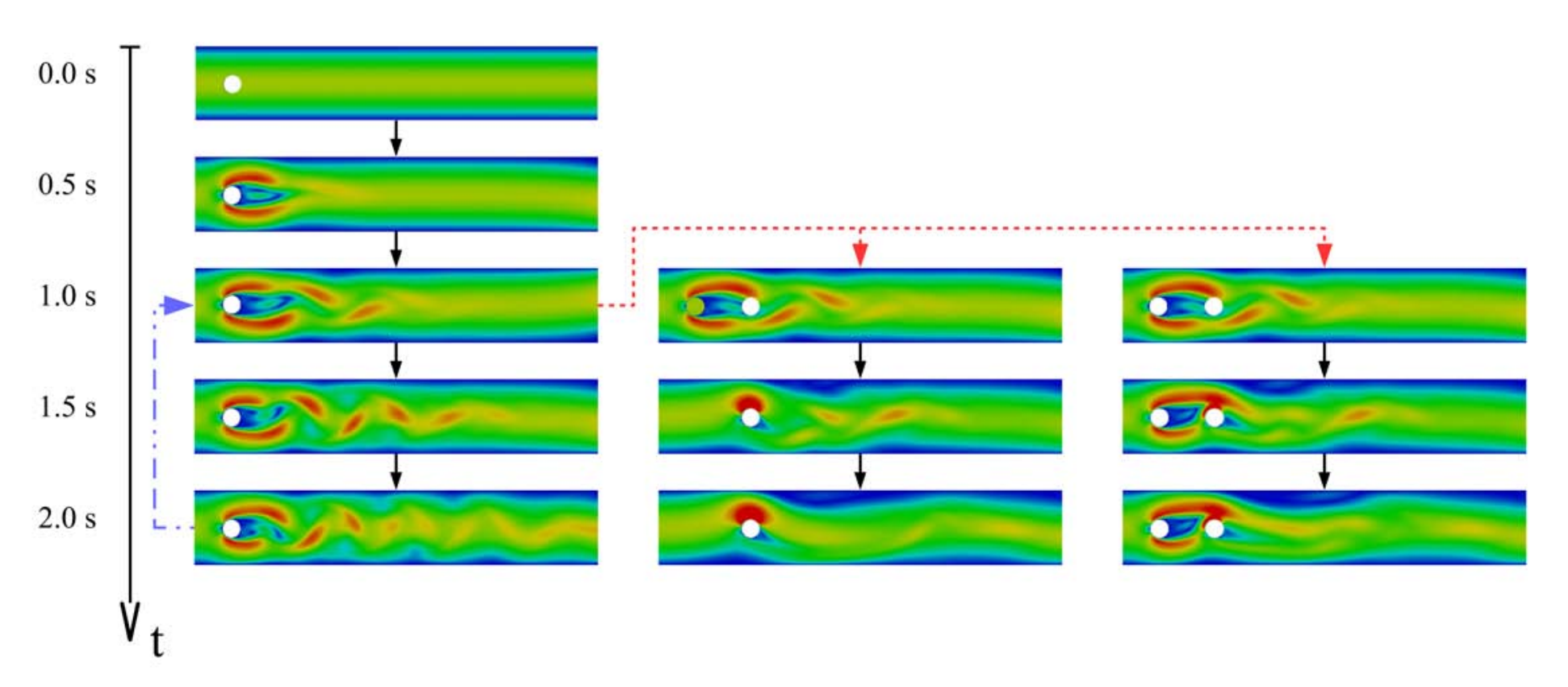}
\caption{Time reversible steering: comparative visualisation of a von K\'{a}rm\'{a}n vortex street according to \cite{bib:SchaeferTurekDurstEtAl1996} at selected times from $t = 0.0$\,s and $t = 2.0$\,s for the basic setup and two altered scenarios -- going back to a previous time step (blue dashed and dotted arrow) and modifying boundary conditions (red dashed arrows) -- restarting at $t = 1.0$\,s}
\label{fig:TRS_Karman}
\end{figure}

The first case is a benchmark scenario initially proposed by Sch\"afer and Turek within the DFG priority research program `Flow Simulation on High-Performance Computers' \cite{bib:SchaeferTurekDurstEtAl1996}. The setup consists of a two dimensional channel flow with a cylinder obstacle near the inlet on the left-hand side channel boundary. Using Reynold's number $Re = 100$, an unsteady flow is generated which leads to the well known phenomena of vortex shedding behind the obstacle. The basic scenario was simulated for two seconds, then, the simulation was rolled back to the one second mark, boundary conditions have been altered and the simulation was continued. Fig.~\ref{fig:TRS_Karman} shows visualisations from $t = 0.0$\,s to $t = 2.0$\,s of the basic setup on the left side. Visualisations from shifting the obstacle at $t = 1.0$\,s are shown in the middle. Finally, introducing a second obstacle at $t = 1.0$\,s is shown on the right. Again it is emphasised that these are not separate simulations, but rather branchings within the framework. 

The second example conducted is a thermally coupled airflow simulation of an operation theatre with a complex geometry. The standard configuration consists of the operation theatre, one patient and two assistants. This example was initially shown in \cite{bib:Wenisch2008} using an isothermal setting and further analysed in \cite{bib:Frisch2014}, this time using a thermally coupled approach. The air inlet is realised over one complete wall, while a slightly open door on the opposite wall functions as an outlet. Temperate boundary conditions are set on all geometry objects, $T = 324.66$\,K on the lamps, $T = 299.50$\,K on human models and $T = 290.16$\,K on all other objects. 

A desirable configuration is signified by an airflow streaming away from the patient, as to minimise the risk of germs entering surgery wounds. This example converges to a steady state, as such, initial time steps are computationally more expensive than latter ones. Therefore, the time reversible steering concept -- restarting from a reasonable converged state, altering the boundary conditions slightly -- proves exceptionally well for quickly iterating through different design possibilities and assessing their applicability, while skipping the expensive initial simulation. 

\begin{figure}[!]
  \begin{subfigure}[b]{\textwidth}
  \centering
    \includegraphics[height=3.5in,width=\textwidth]{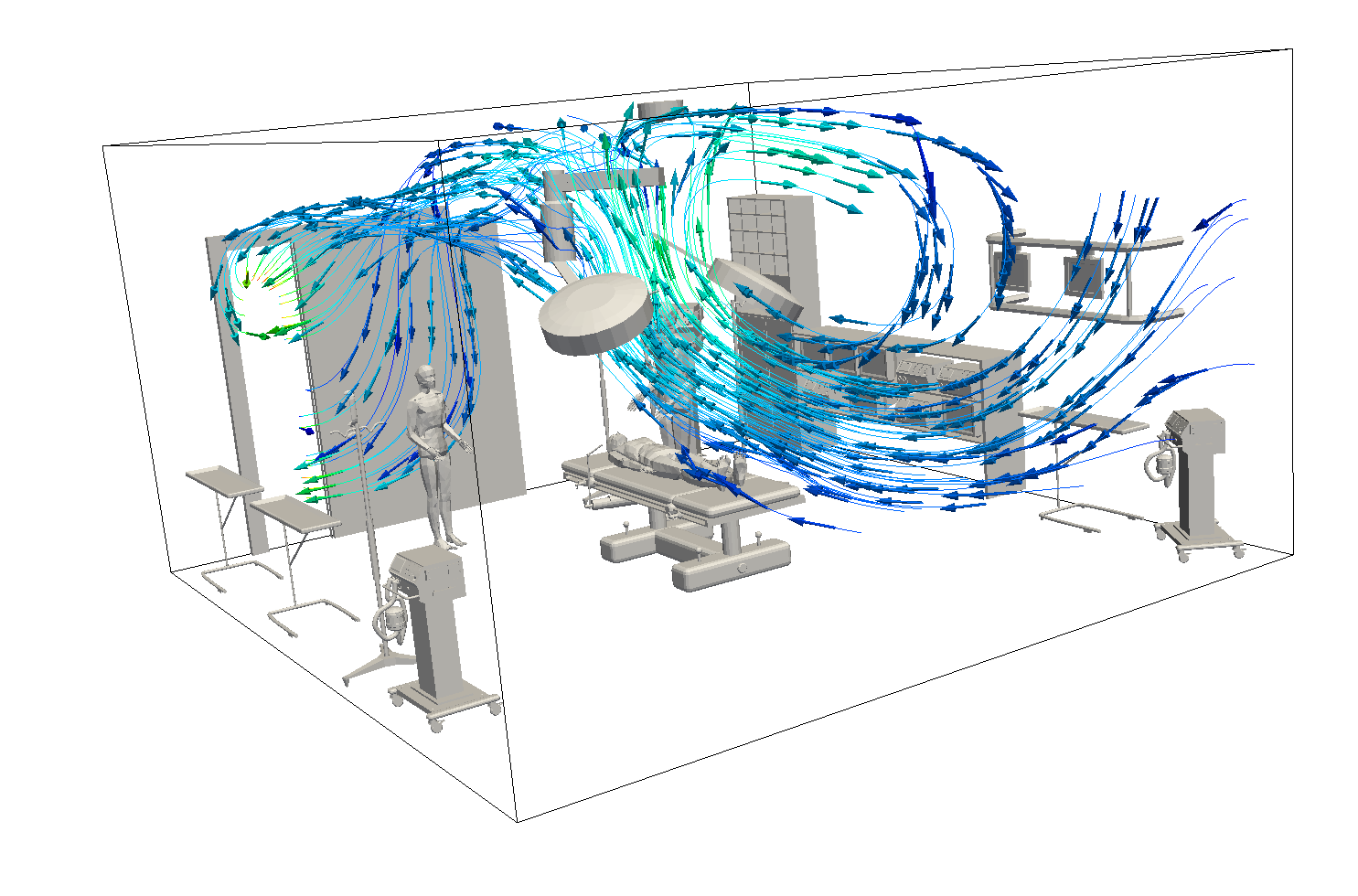}
    \caption{Operation theatre at $t = 50.0$\,s with lamps at $T = 324.66$\,K}
    \label{fig:OP_2A}
  \end{subfigure}
    \hfill
    \begin{subfigure}[b]{\textwidth}
    \centering
    \includegraphics[height=3.5in,width=\textwidth]{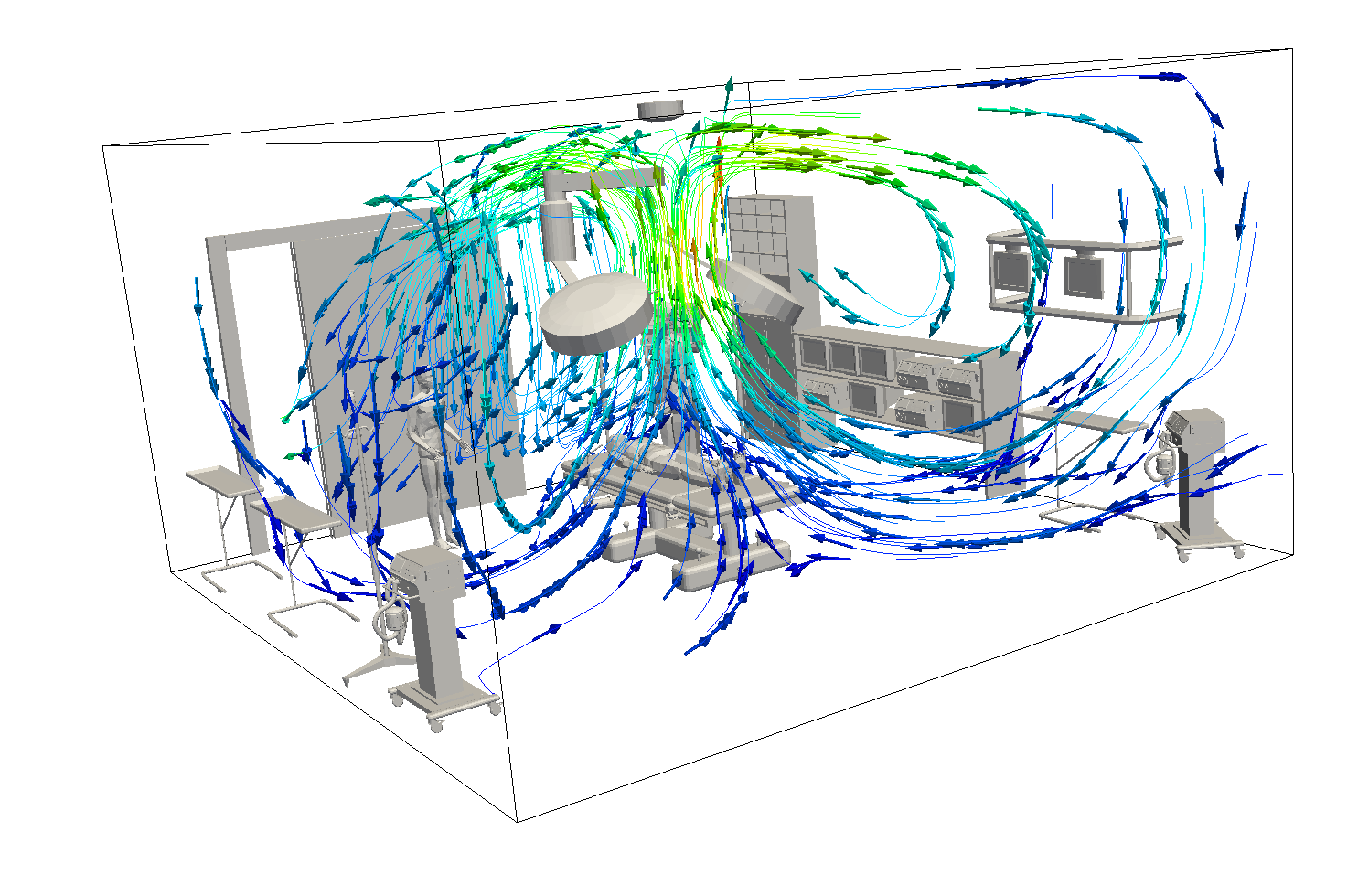}
    \caption{Operation theatre at $t = 50.0$\,s with lamp boundary conditions altered at $t = 20.0$\,s  to $T = 374.66$\,K}
    \label{fig:OP_2AH}
  \end{subfigure}
  \caption{Simulation of an Operation theatre, utilising the TRS concept to evaluate two different temperature boundary conditions applied on the lamps.}
\end{figure}

The visualisations in Fig. \ref{fig:OP_2A} and Fig. \ref{fig:OP_2AH} depict the outcome of converged states at elapsed time $t = 50.0$\,s with two different temperature boundary conditions applied to the lamps. The first scenario is simulated for the full $50.0$\,s, reloaded at elapsed time $t = 20.0$\,s, then, the temperature is increased by $50$\,°K and simulation is resumed. On the author's in-house cluster, an Intel Sandybridge architecture using 16 cores, the first $20.0$\,s of simulation take roughly $24$\,h, while the later $30.0$\,s take $12$\,h. Using TRS, one is able to evaluate the altered state at approximately $33\%$ of time investment compared to a full simulation run. 

\section{Output Performance}
\label{sec:PerfMeasurements}
The following section highlights the performance measurements of the kernel's write routines, the most performance critical aspect of the implementation as these are carried out continuously during a simulation run. Extensive measurements were conducted on the JuQueen supercomputer located at J\"ulich Supercomputing Centre. Furthermore, the kernel was recently deployed on SuperMUC and initial testing results are presented in the final part of this section.

\subsection{Benchmark Systems}
\label{subsec:Systems}
The JuQueen is an IBM BlueGene/Q system which combines 27,672 computing nodes and 458,752 cores in 28 racks. Each node employs a memory of 16\,GB, amounting to 448\,TB for the whole installation. Intra-rack communication is realised via a five dimensional torus network, made up from high speed serial links. The system's theoretical peak performance is listed at 5.9\,Petaflops, with a sustained Linpack performance at 5.0\,Petaflops. For I/O, the system employs dedicated I/O nodes, grouped in I/O drawers installed in the top compartment of the racks. Each rack except the last one contains one I/O drawer with eight I/O nodes. Every I/O node has two PCIe ports for input and output connecting to the torus network, allowing 4\,GB/s of raw data throughput one way. This sums up to a bandwidth of 32\,GB/s per rack. Each I/O node is connected via two 10GbE Ethernet adapters to the file system, allowing a maximal bandwidth of 16\,GB/s per I/O drawer. The underlying parallel file system is IBM's General Parallel File System (GPFS). A more comprehensive overview of the the JuQueen can be found in the official Documentation \cite{bib:JUQ2015} and the best practice guide from PRACE \cite{bib:WauteletBoiarciucDupaysEtAl2014}.

The second system, SuperMUC, employs 18 so called Thin node Islands, each consisting of 512 System x iDataPlex dx360 M4 compute nodes. With two Sandy Bridge-EP Xeon E5-2680 8C processors per node and eight cores per processor the complete system combines 147,456 computing cores. SuperMUC utilises a non-blocking tree intra-island and a 4:1 pruned tree inter-island network topology using Infiniband FD10 switches. The underlying file system is GPFS, similar to JuQueen, with a combined bandwidth of all islands of 200\,GB/s. More detailed information on SuperMUC can be found in its respective best practice guide from PRACE \cite{bib:AnastopoulosNikunenWeinberg2013}.

\subsection{Hardware-Aware Optimisations}
\label{subsec:HardwareOp}

To achieve optimised performance, the I/O kernel has to be aware of the underlying hardware and must be tuned accordingly. Most of the optimisations like alignment of data to the file system's block size lead only to comparably small improvements in write speed, however enabling collective buffering and disabling expensive file locking mechanisms of the GPFS have proven to be indispensable for the performance of the kernel.

Collective buffering utilises a subset of the computing nodes as aggregators, which collect data from the different processes and manage the file accesses. The JuQueen's node cabling is qualified exceptionally well for collective buffering. Each computing node employs 10 links to the intra-rack five dimensional torus network. However, only 16 of the 1024 nodes employ a single link to the I/O nodes. Doing independent I/O over the very scarce amount of I/O connections would lead to severe contention and minuscule performance. The natural choice for the aggregators are the nodes that employ the direct links to the I/O drawers. Data is collected over the very fast intra-rack network while the I/O links are utilised to their full extent. 

To avoid contention by concurrent file accesses, the file driver of MPI-IO's current implementation on the JuQueen employs a very conservative file locking policy which proves detrimental to the performance of shared file approaches. Since each participating rank has its exclusive access region, it is safe to disable the file locking, thus leading to a tremendous increase in performance.

\subsection{Measurements and Results}
\label{subsec:Measurements}

\begin{figure}[!t]
  \centering
  \begin{subfigure}[b]{0.7\textwidth}
    \includegraphics[width=\textwidth]{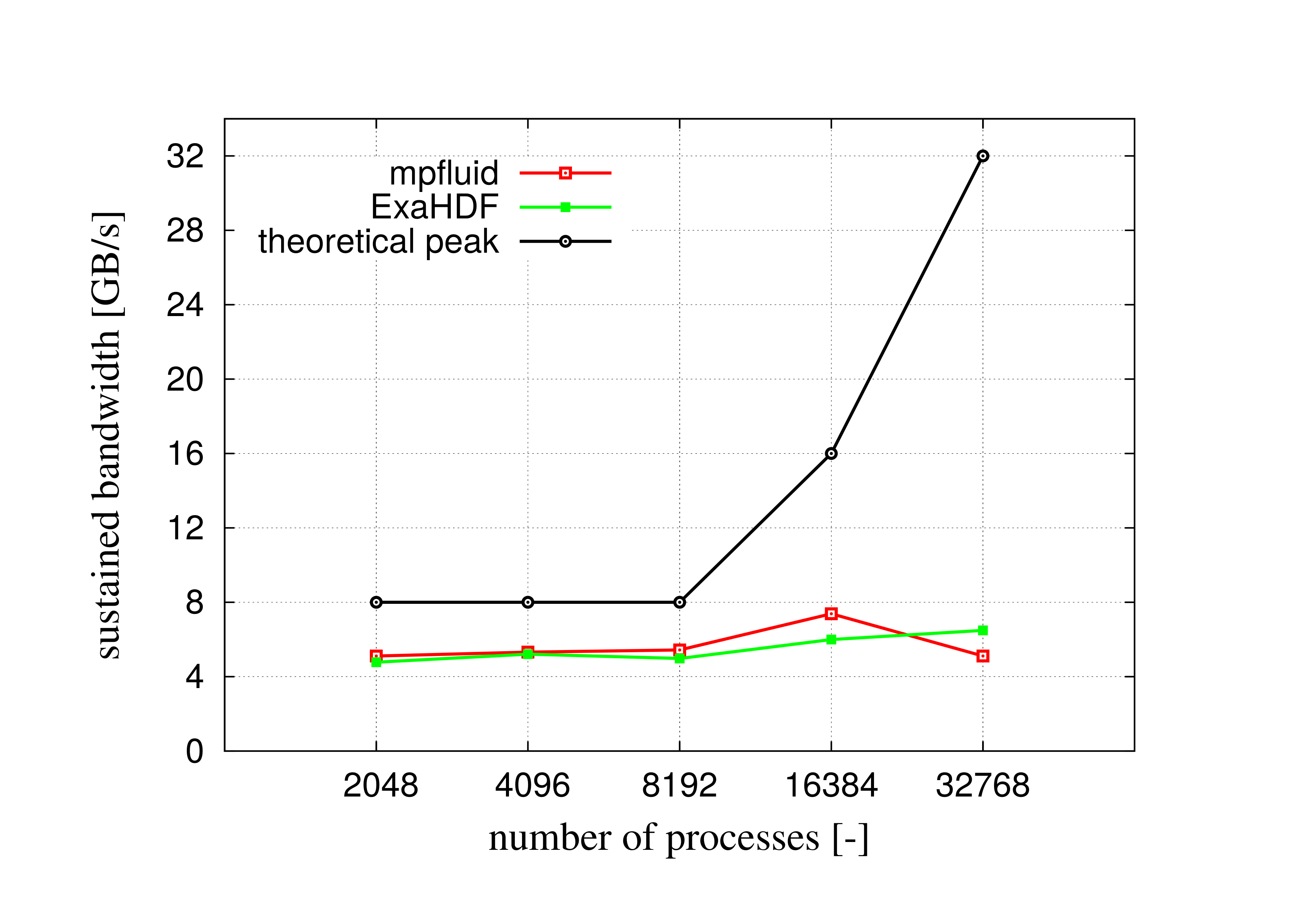}
    \caption{Fully refined 3D domain of resolution $1024 \times 1024 \times 1024$ with approx.\ 11 billion unknowns and a total file size of 337\,GB}
    \label{fig:scaling1}
  \end{subfigure}
  \hfill
    \centering
  \begin{subfigure}[b]{0.7\textwidth}
    \includegraphics[width=\textwidth]{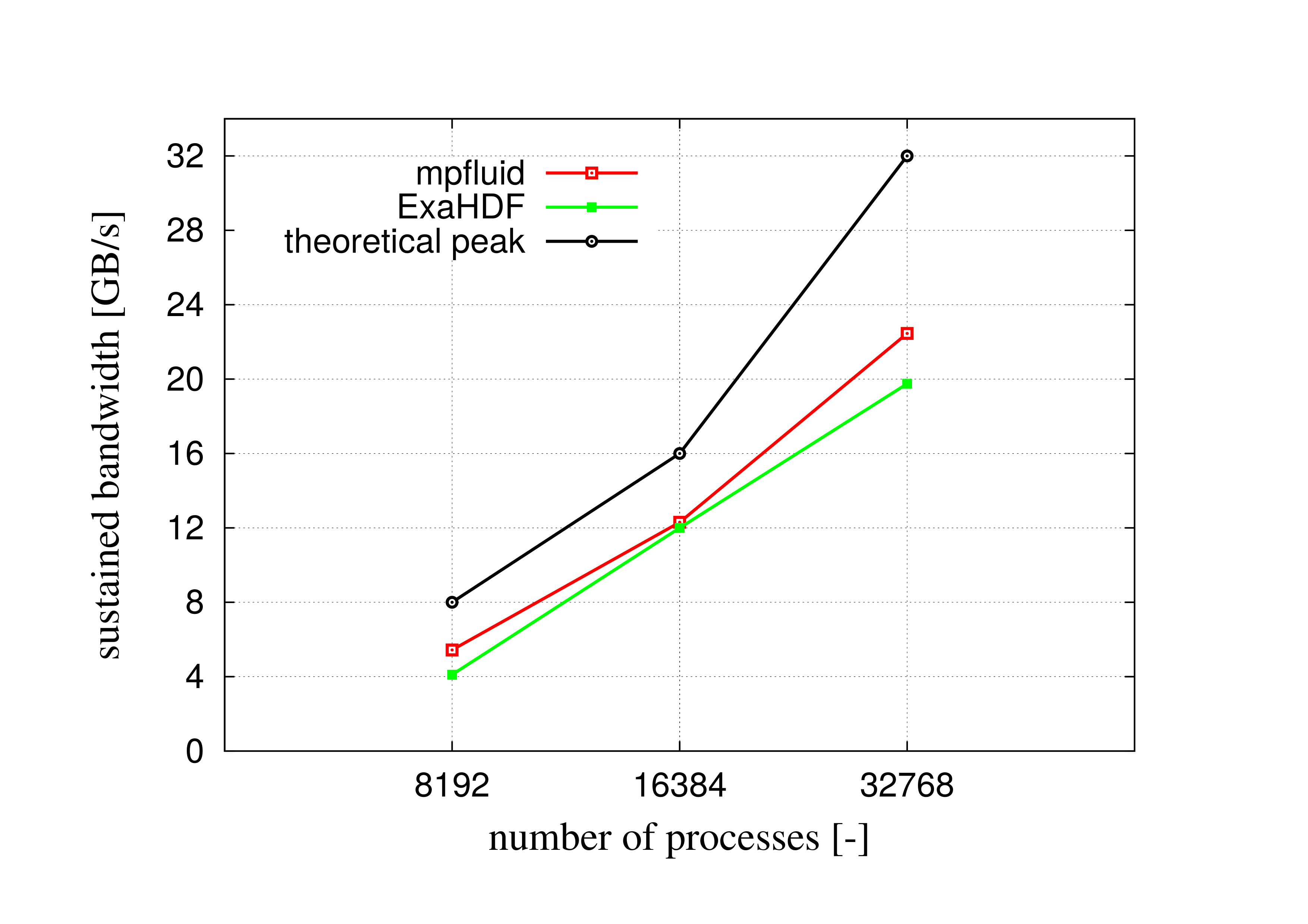}
    \caption{Fully refined 3D domain of resolution $2048 \times 2048 \times 2048$ with approx.\ 90 billion unknowns and a total file size of 2,750\,GB}
    \label{fig:scaling2}
  \end{subfigure}
  \caption{Sustained bandwidth of a write out for two domain configurations plotted against different amounts of processes on the JuQueen}
\end{figure}

To be able to classify our CFD code's I/O performance not only in terms of utilised bandwidth, \emph{mpfluid}'s kernel was compared against another I/O kernel based on HDF5. As reference, the VPIC-IO (vector particle-in-cell) kernel \cite{bib:BynaHowison2015} from ExaHDF used in the largest I/O run to date \cite{bib:BynaUseltonPrabhatEtAl2013} was employed. In order to get architecture independent measurements, comparable to the present implementation, the measurements using VPIC-IO were done similarly on the JuQueen, using the same optimisations and scaling the total amount of data for both kernels to be equal. 

Two test cases were used, varying the amount of computing processes utilised. The first test case involved a fully refined 3D domain of resolution $1024 \times 1024 \times 1024$ (depth 6) leading to a total amount of about 300,000 d-grids in the domain. Each d-grid contains 16 cells in every dimension, making up 4096 cells per single d-grid and about 1.23 billion cells in the entire domain. Each written checkpoint claims a file size of 337\,GB. 

The second test case uses the same properties but resolves the domain one level further up to a resolution of $2048 \times 2048 \times 2048$ (depth 7), resulting in a total amount of about 2.4 million d-grids and 10 billion cells at a checkpoint size of 2.7\,TB. 

The refinement levels as well as the amount of cells per computational grid were taken in accordance to Frisch and Mundani \cite{bib:FrischMundani15}, who determined the setup to perform very well for solving the underlying Navier Stokes Equations.

Fig.~\ref{fig:scaling1} shows excellent performance for both kernels in the range from 2048 to 8192 processes. The
similar measurements are attributable to the equal I/O resources they have available. If no other application is using the I/O nodes, all four nodes, connected to one half of the rack are available within the intra-rack network. The discrepancy between the measured and the theoretical peak bandwidth is believed to be due to the wind up and wind down of write operations to individual datasets. Here more in-depth measurements are required. Using 16,384 processes and having a full drawer with eight I/O nodes available the used bandwidth increased by about 20\,\%. An unsatisfactory result, as now double the I/O resources are available compared to the previous cases. Doubling the amount of processes and available I/O nodes again, reveals even worse scaling as only one forth of the estimated bandwidth compared to the first three measurements was achieved. The authors believe this is attributable to the amount of grids per process on the later test cases. While each individual process has fewer grids to manage, the communication overhead of filling the aggregators' write buffers increases, which is likely to be responsible for the bad scaling behaviour. Additionally, more aggregators, assuming less data each participate in I/O, certainly affects the performance negatively. On top of the fact that the problem is too small to show adequate scaling in the regions above 8192 processes for I/O performance, the same was observed for the actual computation in \cite{bib:FrischMundani15}.

For the second test case, measurements were not possible below 8192 processes due to the aforementioned memory limitation introduced by the additional write buffers. Fig.~\ref{fig:scaling2} shows the measured results. As expected, the measurements show adequate scaling in the expected range, both for VPIC-IO's and our kernel. 

The same test scenario as was run on JuQueen was performed on SuperMUC, with a fully refined 3D domain resolved until depth 6, amounting to approx. 1.23 billion cells and a checkpoint size of 337\,GB. The best result was achieved using 2048 processes at 21.4\,GB/s with a gradual decrease in performance using 4096 processes at 14.92\,GB/s and 8192 processes at 4,64\,GB/s. This decrease in performance is in accordance with the findings gained from JuQueen, stating that below a certain amount of grids per process communication overhead becomes an issue. The higher bandwidth at a lower node count in comparison to the JuQueen is attributable to the different network topology, which in case of the SuperMUC does not exhibit the aforementioned I/O bottleneck.

\section{Conclusion}
In this paper, a massive parallel CFD code was presented that was successfully deployed on two of Germany's supercomputers, running on about 140,000 cores, solving a problem with more than 700 billion unknowns. Core of this work is a hierarchical data structure that not only supports distributed computing and the development of efficient numerical solvers, but further allows users an in-situ visualisation (sliding window) of the computed results in order to leverage high-performance interactive data exploration. This code was now extended by an HDF5 I/O kernel to tackle the problem of reading and writing huge datasets during runtime, a typical bottleneck for modern HPC applications, thus slowing down the overall performance due to long I/O latencies.

The current I/O kernel -- using collective buffering for optimisation -- shows a good scaling behaviour within the applicable range of grids, that is, any limiting factor is due to the specifically used I/O hardware, so that with an increasing amount of I/O nodes a much better performance is to be expected. Further comparisons with ExaHDF's I/O kernel foster these results. By adopting above data structure for the design of the kernel's internal file structure, additional functionality such as a file-based sliding window could easily be implemented, now providing users the opportunity for a time reversal steering, enabling to change boundary conditions and restart the computation at any previous time step.

\section{Acknowledgement}

The authors gratefully acknowledge the computing time granted by the JARA-HPC Vergabegremium and provided on the JARA-HPC Partition part of the supercomputer JUQUEEN \cite{Juqueen2015} at Forschungszentrum J\"ulich. Futhermore, the authors would like to cordially thank Leibniz Supercomputing Centre (LRZ) for providing computing time on SuperMUC. Without their kind support, parts of this work would not have been possible.

\bibliographystyle{IEEEtran}
\bibliography{paper}

\begin{thebibliography}{10}
\providecommand{\url}[1]{#1}
\csname url@samestyle\endcsname
\providecommand{\newblock}{\relax}
\providecommand{\bibinfo}[2]{#2}
\providecommand{\BIBentrySTDinterwordspacing}{\spaceskip=0pt\relax}
\providecommand{\BIBentryALTinterwordstretchfactor}{4}
\providecommand{\BIBentryALTinterwordspacing}{\spaceskip=\fontdimen2\font plus
\BIBentryALTinterwordstretchfactor\fontdimen3\font minus
  \fontdimen4\font\relax}
\providecommand{\BIBforeignlanguage}[2]{{%
\expandafter\ifx\csname l@#1\endcsname\relax
\typeout{** WARNING: IEEEtran.bst: No hyphenation pattern has been}%
\typeout{** loaded for the language `#1'. Using the pattern for}%
\typeout{** the default language instead.}%
\else
\language=\csname l@#1\endcsname
\fi
#2}}
\providecommand{\BIBdecl}{\relax}
\BIBdecl

\bibitem{bib:kalilmiller2015}
\BIBentryALTinterwordspacing
T.~Kalil and J.~Miller. (2015) Advancing {U.S}. leadership in high-performance
  computing. [Online]. Available:
  \url{www.whitehouse.gov/blog/2015/07/29/advancing-us-leadership-high-performance-computing}
\BIBentrySTDinterwordspacing

\bibitem{bib:babufrmu2002}
M.~Bader, H.-J. Bungartz, A.~Frank, and R.-P. Mundani, ``Space tree structures
  for {PDE} software,'' in \emph{Computational Science}, ser. LNCS 2331,
  P.~Sloot, C.~Tan, J.~Dongarra, and A.~Hoekstra, Eds.\hskip 1em plus 0.5em
  minus 0.4em\relax Springer, 2002, pp. 662--671.

\bibitem{bib:mufrvara2015}
R.-P. Mundani, J.~Frisch, V.~Varduhn, and E.~Rank, ``A sliding window technique
  for interactive high-performance computing scenarios,'' \emph{Advances in
  Engineering Software}, vol.~84, pp. 21--30, 2015.

\bibitem{bib:ricamusl2012}
M.~Rivi, L.~Calori, G.~Muscianisi, and V.~Slavnic, ``In-situ visualization:
  {S}tate-of-the-art and some use cases,'' in \emph{PRACE White Paper}.\hskip
  1em plus 0.5em minus 0.4em\relax Belgium: PRACE Brussels, 2012.

\bibitem{bib:erfrmu2016}
C.~Ertl, J.~Frisch, and R.-P. Mundani, ``Massive parallel fluid flow
  simulations using hierarchical data format version 5 ({HDF5}),'' in
  \emph{Proc. of the 15th Int. Symposium on Parallel and Distributed
  Computing}.\hskip 1em plus 0.5em minus 0.4em\relax IEEE Computer Society,
  2016, pp. 30--37.

\bibitem{bib:frmuratr2015}
J.~Frisch, R.-P. Mundani, E.~Rank, and C.~{van~Treek}, ``Engineering-based
  thermal {CFD} simulations on massive parallel systems,'' \emph{computation},
  no.~3, pp. 235--261, 2015.

\bibitem{bib:ferzperic2002}
J.~Ferziger and M.~Peri\'c, \emph{Computational Methods for Fluid Dynamics},
  3rd~ed.\hskip 1em plus 0.5em minus 0.4em\relax Springer, 2002.

\bibitem{bib:hirsch2007}
C.~Hirsch, \emph{Numerical Computation of Internal and External Flows},
  2nd~ed.\hskip 1em plus 0.5em minus 0.4em\relax Butterworth--Heinemann, 2007,
  vol.~1.

\bibitem{bib:lienhard2011}
J.~{Lienhard~IV} and J.~{Lienhard~V}, \emph{A Heat Transfer Textbook}, 4th~ed.,
  ser. Dover Civil and Mechanical Engineering.\hskip 1em plus 0.5em minus
  0.4em\relax Dover Publications, 2011.

\bibitem{bib:harlwelch1965}
F.~Harlow and J.~Welch, ``Numerical calculation of time-dependent viscous
  incompressible flow of fluid with free surface,'' \emph{Physics of Fluids},
  vol.~8, no.~12, pp. 2182--2189, 1965.

\bibitem{bib:chorin1968}
A.~Chorin, ``Numerical solution of the navier-stokes equations,''
  \emph{Mathematics of Computation}, vol.~22, no. 104, pp. 745--762, 1968.

\bibitem{bib:frmura2011}
J.~Frisch, R.-P. Mundani, and E.~Rank, ``Communication schemes of a parallel
  fluid solver for multi-scale environmental simulations,'' in \emph{Proc. of
  the 13th Int. Symposium on Symbolic and Numeric Algorithms for Scientific
  Computing}.\hskip 1em plus 0.5em minus 0.4em\relax IEEE Computer Society,
  2011, pp. 391--397.

\bibitem{bib:Frisch2014}
J.~Frisch, ``Towards massive parallel fluid flow simulations in computational
  engineering,'' Ph.D. dissertation, Technische Universit\"at M\"unchen, 2014.

\bibitem{bib:brandt1977}
A.~Brandt, ``Multi-level adaptive solutions to boundary-value problems,''
  \emph{Mathematics of Computation}, vol.~31, no. 138, pp. 333--390, 1977.

\bibitem{bib:frmura2013}
J.~Frisch, R.-P. Mundani, and E.~Rank, ``Parallel multi-grid like solver for
  the pressure {P}oisson equation in fluid flow applications,'' in \emph{Proc.
  of the IADIS Int. Conf. on Applied Computing}.\hskip 1em plus 0.5em minus
  0.4em\relax IADIS Press, 2013, pp. 139--146.

\bibitem{bib:paraview}
\BIBentryALTinterwordspacing
{ParaView}. [Online]. Available: \url{www.paraview.org}
\BIBentrySTDinterwordspacing

\bibitem{bib:HDF5}
\BIBentryALTinterwordspacing
{The {HDF} {Group}}. {{HDF5} {Software} {Documentation}}. {Release} 1.8.16.
  [Online]. Available: \url{www.hdfgroup.org/HDF5/doc/}
\BIBentrySTDinterwordspacing

\bibitem{bib:SchaeferTurekDurstEtAl1996}
M.~Sch{\"a}fer, S.~Turek, F.~Durst, E.~Krause, and R.~Rannacher, ``Benchmark
  computations of laminar flow around a cylinder,'' in \emph{Flow Simulation
  with High-Performance Computers II}, ser. Notes on Numerical Fluid Mechanics,
  vol.~52.\hskip 1em plus 0.5em minus 0.4em\relax Vieweg, 1996, pp. 547--566.

\bibitem{bib:Wenisch2008}
P.~Wenisch, ``Computational steering of {CFD} simulations on
  teraflop-supercomputers,'' Ph.D. dissertation, Technische Universit\"at
  M\"unchen, 2008.

\bibitem{bib:JUQ2015}
\BIBentryALTinterwordspacing
{{JuQueen} -- {Documentation}}. [Online]. Available:
  \url{www.fz-juelich.de/ias/jsc/EN/Expertise/Supercomputers/JUQUEEN/Documentation/Documention_node.html}
\BIBentrySTDinterwordspacing

\bibitem{bib:WauteletBoiarciucDupaysEtAl2014}
P.~Wautelet, M.~Boiarciuc, J.-M. Dupays, S.~Giuliani, M.~Guarrasi,
  G.~Muscianisi, and M.~Cytowski, \emph{Best {Practice} {Guide} - {Blue}
  {Gene/Q} v1.1.1}, PRACE - Partnership for Advanced Computing in Europe, 2014.

\bibitem{bib:AnastopoulosNikunenWeinberg2013}
N.~Anastopoulos, P.~Nikunen, and V.~Weinberg, \emph{Best {Practice} {Guide} -
  {SuperMUC} v1.0}, PRACE - Partnership for Advanced Computing in Europe, 2013.

\bibitem{bib:BynaHowison2015}
\BIBentryALTinterwordspacing
S.~Byna and M.~Howison. (2015) {Parallel {I/O} Kernel ({PIOK}) Suite}.
  [Online]. Available: \url{https://sdm.lbl.gov/exahdf5/software.html}
\BIBentrySTDinterwordspacing

\bibitem{bib:BynaUseltonPrabhatEtAl2013}
S.~Byna, A.~Uselton, Prabhat, D.~Knaak, and Y.~He, ``Trillion particles,
  120,000 cores, and 350 {TBs}: Lessons learned from a {Hero} {I/O} run on
  {Hopper},'' in \emph{Cray User Group meeting}, 2013.

\bibitem{bib:FrischMundani15}
R.-P. Mundani and J.~Frisch, ``{Measuring and comparing the scaling behaviour
  of a high-performance CFD code on different supercomputing
  infrastructures},'' in \emph{{Proc. of the 17th Int. Symposium on Symbolic
  and Numeric Algorithms for Scientific Computing}}.\hskip 1em plus 0.5em minus
  0.4em\relax {{IEEE} Press}, {2015}.

\bibitem{Juqueen2015}
M.~Stephan and J.~Docter, ``{JUQUEEN: IBM Blue Gene/Q {\textcircled{c}}
  Supercomputer System at the J\"ulich Supercomputing Centre},'' \emph{Journal
  of large-scale research facilities JLSRF}, vol.~1, 2015.

\end{thebibliography}

\end{document}